\documentclass[aps,twocolumn]{revtex4}
\usepackage[dvips]{graphics,graphicx}
\begin{document}
\title{Bright solitons and self-trapping with a BEC of cold atoms in driven tilted optical lattices}
\author{Andrey R. Kolovsky}
\affiliation{Kirensky Institute of Physics and Siberian Federal University, 660036 Krasnoyarsk, Russia}

\begin{abstract}
We suggest a method for creating bright matter solitons by loading a BEC of atoms in a driven tilted optical lattice.  It is shown that one can realize the self-focussing regime for the wave-packet dynamics by properly adjusting the phase of the driving field with respect to the phase of Bloch oscillations. If atom-atom interactions are larger than some critical value $g_{min}$, this self-focussing regime is followed by the formation of bright solitons. Increasing the interactions above another critical value $g_{max}$ makes this process unstable. Instead of soliton formation one now meets the phenomenon of incoherent self-trapping. In this regime a fraction of atoms is trapped in incoherent localized wave-packets, while the remaining atoms spread ballistically. 
\end{abstract}

\maketitle

{\em Introduction.}
Non-spreading localized wave-packets, solitons, are a paradigm of nonlinear wave dynamics and are encountered in many different fields, such as physics, biology, oceanography, and telecommunication \cite{book}. Recently much attention has been paid to solitons with a Bose-Einstein condensate (BEC) of cold atoms \cite{Khay02,Stre02,Burg99,Dens00,Blud09,Eier04}, described in the mean-field approximation by the nonlinear Schr\"odinder equation.  To generate solitons this equation should be self-focussing, which requires either attractive inter-atomic interactions or negative mass. The latter case can be achieved by loading a BEC into an optical lattice, where the effective atomic mass 
is negative near the edge of the Brillouin zone. This idea was realized in experiment \cite{Eier04} where a wave-packet of cold atoms was moved to the edge of the zone by accelerating the optical lattice for a given time. Adjusting the number of atoms in the wave-packet and its width the authors of the cited experiment were able to create a stationary soliton, which showed no sign of dispersion for at least 60ms.

In this work we suggest a different method for creating bright solitons by loading a BEC of atoms in a driven tilted optical lattice. This method relies on the well-known result that the quasi-energy spectrum of the system has band structure if the driving frequency coincides with the Bloch frequency (see for example review \cite{PhysRep} and references therein). Changing the phase of the driving field with respect to the phase of Bloch oscillations it is possible to realize both defocussing and focussing regimes of the nonlinear Schr\"odinder equation. We demonstrate that the suggested method is robust and does not require the simultaneous  adjusting of wave-packet width and atom number. Moreover, it allows us to generate moving solitons, which is a challenging problem from the view point of laboratory experiments. 

\bigskip
\noindent
{\em The model.}
Considering deep enough optical lattices, the BEC dynamics in a tilted driven lattice obeys the following equation
\begin{equation}
\label{1}
i\hbar\dot{a}_l= -\frac{J}{2}(a_{l+1}+a_{l-1}) +d[F-F_\omega\cos(\omega t+\phi)] l a_l +g|a_l|^2 a_l \;,
\end{equation}
where $a_l$ is the BEC complex amplitude in $l$th potential well, $J$ the hopping matrix element, $d$ the lattice period, $F$ and $F_\omega$ magnitudes of a static and AC fields, $\omega$ the driving frequency, and $\phi$ an arbitrary phase. For vanishing atom-atom interactions system (\ref{1}) has been discussed in numerous papers, including the experimental works \cite{Sias08,Ivan08,Albe09}.  For $g\ne0$ and arbitrary values of the driving frequency system (\ref{1}) has been studied in the recent experimental work \cite{communic} and the theoretical work \cite{preprint}. Here we focus on the particular case where the driving frequency $\omega$ coincides with the Bloch frequency $\omega_B=dF/\hbar$. As initial conditions we consider a coherent wave-packet of a finite width and, to be specific, we assume $a_l(t=0)\sim\exp( -l^2/2\sigma_0^2)$. 

One gets a useful insight into the BEC dynamics by changing from the Wannier basis to the basis of Wannier-Stark states, which corresponds to the substitution
\begin{equation}
\label{3}
c_m=\sum_l {\cal J}_{m-l}(J/dF) a_l \;.
\end{equation}
Assuming additionally $F_\omega\ll F$  and using the rotating-wave approximation, the equation with new amplitudes $c_m$  takes the form
\begin{displaymath}
i\hbar\dot{c}_m= -\frac{\widetilde{J}}{2}(c_{m+1}e^{i\phi}+c_{m-1}e^{-i\phi}) 
\end{displaymath}
\begin{equation}
\label{4}
+ g\sum_{m_1,m_2,m_3} I_{m,m_1,m_2,m_3} c^*_{m_1}c_{m_2}c_{m_3}\delta(m+m_1-m_2-m_3)  \;,
\end{equation}
where  $\widetilde{J}=JF_\omega/2F$ and the kernel $I_{m,m_1,m_2,m_3}=\sum_n {\cal J}_{n-m}(z) {\cal J}_{n-m_1}(z) {\cal J}_{n-m_2}(z) {\cal J}_{n-m_3}(z)$. (Here and above ${\cal J}_{n}(z)$ are Bessel functions of the first kind and $z=J/dF$.) Since we have assumed $\omega=\omega_B$, the parameter $\phi$ in Eq.~(\ref{4}) takes into account the phase difference between the Bloch and field oscillations.
 
{\em Dynamics for $dF\gg J$.}
We begin our analysis of  Eq.~(\ref{4}) by considering the case of a strong static field, $dF\gg J$. In this limiting case the Wannier-Stark states coincide with the Wannier states and Eq.~(\ref{4}) reduces to the celebrated Discrete Nonlinear Schr\"odinger Equation 
\begin{equation}
\label{6}
i\hbar\dot{c}_m= -\frac{\widetilde{J}}{2}(c_{m+1}e^{i\phi}+c_{m-1}e^{-i\phi})  + g|c_m|^2 c_m  \;.
\end{equation}
This equation can be considered as a discretization of the continuous nonlinear  Schr\"odinger equation
\begin{equation}
\label{7}
i\hbar\frac{\partial \psi}{\partial t}=-\frac{\widetilde{J}}{2} 
\left[d^2\cos\phi\frac{\partial^2 \psi}{\partial x^2}
+2id\sin\phi\frac{\partial \psi}{\partial x} \right]+g|\psi|^2\psi \;,
\end{equation}
where $x=md$ and the irrelevant term $\widetilde{J}\cos\phi \; \psi$ is omitted.

Let us discuss the system (\ref{6}) in terms of its continuous counterpart (\ref{7}). The dynamical regimes of (\ref{7}) depend on the value of the parameter $\phi$. The most interesting cases are $\phi=0$ (defocussing of the initial wave-packet), $\phi=\pm\pi/2$ (translation), and $\phi=\pi$ (self-focussing). It is easy to check numerically that under the condition $\sigma_0\gg 1$ these regimes are present in the discrete equation (\ref{6}) as well. From now on we will focus on the case $\phi=\pi$ where the effective hopping matrix element is negative and Eq.~(\ref{7}) is self-focussing.  As known in the 1D nonlinear Schr\"odinger equation the transient focussing regime is followed by formation of several solitons moving with different velocities \cite{Zakh72}.  In the co-moving frame each of isolated solitons is approximated by the one-parameter function,
\begin{equation}
\label{8}
f(x;\lambda)=\sqrt{\frac{\widetilde{J}}{g}}\frac{\lambda}{\cosh(\lambda x)} \;,
\end{equation}
which solves the stationary nonlinear Schr\"odinger equation. The parameter $\lambda$ in (\ref{8}), which defines the soliton width, is proportional to the number of atoms captured by the soliton and the nonlinearity $g$. Namely, assuming the normalization of the wave function $\psi(x,t)$ to unity,  we have $\lambda=\alpha g/2\widetilde{J} d$, where $\alpha$ is the relative number of atoms in a given soliton. Simulating the dynamics of a single (moving) soliton one finds the function (\ref{8}) with discrete $x=md$ to be an approximate solution of (\ref{6}) until $\lambda\approx0.1$, where the soliton width is about 20 lattice periods. For a larger $\lambda$ the discreetness of $x$ becomes important and one speaks about a discrete soliton or breather \cite{Kivs93,Flach04}. The discrete soliton may involve only 3-4 sites and is pinned in the lattice. 
\begin{figure}
\center
\includegraphics[width=8.5cm]{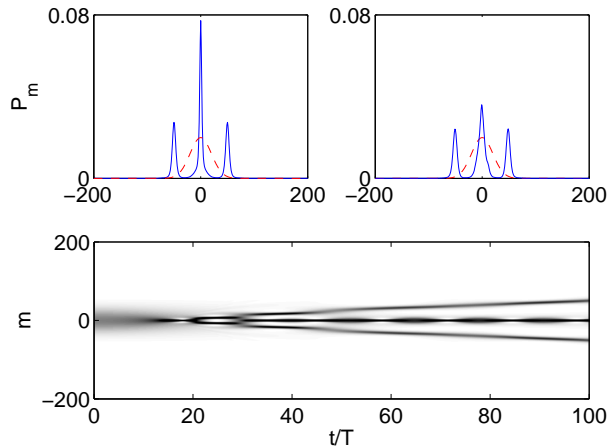}
\caption{Lower panel: Dynamics of (\ref{6}) for $\phi=\pi$,  $\sigma_0=20$, $\widetilde{J}=1$, and $g=2$. The time is measured in units of the tunneling period $T=2\pi\hbar/\widetilde{J}$, which defines the characteristic time scale of the system dynamics. The upper limit of the color axis for the gray-scaled encoding of site populations is set to 0.04.  Upper panels:  Site populations $P_m=|c_m|^2$ at the end of numerical simulations for $N=\infty$ (left) and $N=10^6$ (right).}
\label{Fig1}
\end{figure}

\begin{figure}
\center
\includegraphics[width=8.5cm]{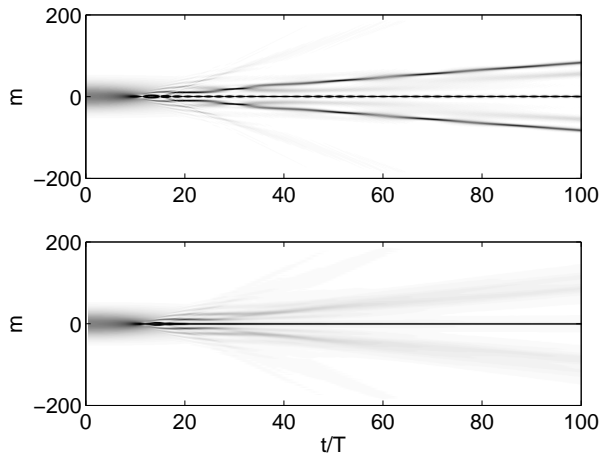}
\caption{The same as in Fig.~\ref{Fig1} yet  $g=4$. The upper and lower panels refer to $N=\infty$  and $N=10^6$, respectively. The central breather in the upper panel captures approx. 40\%  of the atoms, the largest side solitons approx. 5\% each. }
\label{Fig2}
\end{figure}

We simulated the DNLSE (\ref{6}) for a wide initial wave packet, searching for stable soliton-like structures, and indeed, observed them in a certain interval of nonlinearity $g_{min}<g<g_{max}$. An example is given in Fig.~\ref{Fig1}, where $\tilde{g}=g/\widetilde{J}=2$. (Since $\widetilde{J}$ in Eq.~(\ref{6}) can be set to unity by rescaling the time, only the ratio  $g/\widetilde{J}$ matters.) In Fig.~\ref{Fig1} one sees two solitons of the form (\ref{8}) moving in opposite directions with constant velocities, and one central soliton, pinned at the lattice origin. This central soliton, which captures the largest part of atoms, is actually on the border of validity of the continuous approximation and carries some features of discrete solitons. Increasing the interaction constant $g$ more atoms become captured in the central soliton-like structure and it becomes a discrete breather with approximately three sites involved in the dynamics (see upper panel in Fig.~\ref{Fig2}). The remaining atoms are emitted in the form of plane waves and tiny solitons. However, these solitons appear to be sensitive to the interaction-induced decoherence process present in the system and, hence, are of little interest. 

Next we discuss the destructive effect of the interaction-induced decoherence in more detail. 
To mimic the quantum decoherence within the framework of the DNLSE the solution of (\ref{1}) should be averaged over an ensemble of initial conditions. This is uniquely defined by the initial quantum many-body state of the system \cite{Grae09}. In the case of the BEC initial state being considered at present, this ensemble can be approximated by the following ensemble,
\begin{equation}
\label{10}
c'_l=e^{i\theta_l}\sqrt{|c_l|^2+\epsilon_l} \;,\quad 
\overline{\epsilon_l^2}=|c_l|^2/N \;,\quad  \overline{\theta_l^2}=1/4N |c_l|^2 \;,
\end{equation}
where  $\epsilon_l$ and $\theta_l$ are normally distributed random variables and $N$ is the total number of atoms.  Clearly, the ensemble (\ref{10}) takes into account quantum fluctuations of atom number in any given well of the optical lattice. For large $N$  and providing the solution of the DNLSE is stable with respect to a small variation in initial conditions, this additional averaging procedure may be omitted. However, if the solution is not stable this averaging procedure is absolutely necessary. Firstly, because it removes artifacts of the mean-field discription \cite{Grae09,Trim08} and secondly, because it makes transparent the decoherence of the BEC in the course of time \cite{remark0}.

The upper panels in Fig. \ref{Fig1} compare the single-run solution of DNLSE with that averaged over the ensemble (\ref{10}), where we set $N=10^6$. It is seen that the averaging procedure only slightly increases the soliton widths; in all other aspects the time evolution of site populations follow that depicted in the lower panel in Fig.~\ref{Fig1}. However, this result holds only if $\tilde{g}<\tilde{g}_{max}$, where the system dynamics are stable ($\tilde{g}_{max}\approx2$ for the chosen parameters).  A comparison of the upper and lower panels in Fig.~\ref{Fig2}, where $\tilde{g}>\tilde{g}_{max}$ and the system dynamics are unstable, indicates that the decoherence process destroys the small side solitons. Simultaneously it transforms the large central breather into an incoherent packet. For this reason we avoid using the term `discrete soliton' or `breather', which is reserved in physical literature for a coherent localized excitation. Instead we shall use the term `self-trapping', which may be used for both coherent and incoherent localized wave-packets. It is also worth mentioning that the discussed (incoherent) self-trapping principally differs from a (coherent) self-trapping in the defocussing regime (positive hopping matrix element) studied in Refs.~\cite{Moli92,Anke04}.  We come back to the problem of incoherent self-trapping in the next section.

\begin{figure}
\center
\includegraphics[width=8.5cm]{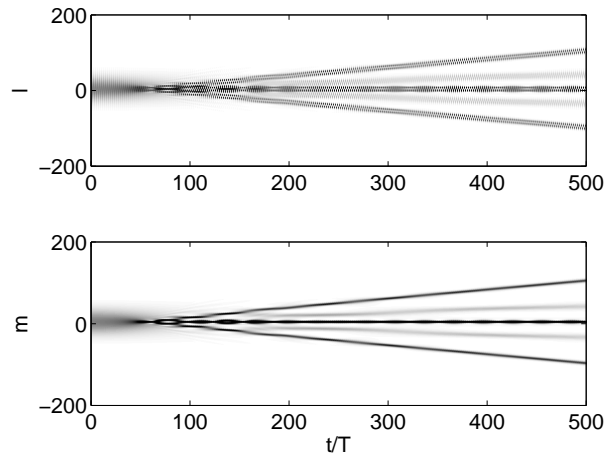}
\caption{Dynamics of the original system (\ref{1}) for $\phi=\pi$,  $\sigma=20$, $N=10^6$, $J=2$, $dF=0.5$, $F_\omega=0.2F$, and  $g=0.8$ in the Wannier (upper panel) and Wannier-Stark (lower panel) basis. The chosen parameters correspond to $\widetilde{J}=0.2$ and $\tilde{g}=4$. The time is measured in units of the effective tunneling period $T=2\pi\hbar/\widetilde{J}$}
\label{Fig3}
\end{figure}

{\em Dynamics for $dF< J$.}
We also studied the formation of bright solitons in the driven tilted lattice (\ref{1}) in the parameter region, where it cannot be reduced to the standard DNLSE. In particular, when relaxing the condition $dF\gg J$ the sum in the nonlinear term in Eq.~(\ref{4}) contains many terms; thus, one cannot appeal to the nonlinear Schr\"odinger equation. Nevertheless, we find the dynamics of system (\ref{4}) to be similar to that of system (\ref{6}).  Namely, for $\phi=0$ one observes defocusing of the initial wave packet, translation dynamics for $\phi=\pi/2$, and self-focusing for $\phi=\pi$. Moreover, the focussing regime is followed by formation of a number of solitons with the characteristic shape given in Eq.~(\ref{8}). An example is given in the lower panel in Fig.~\ref{Fig3}, where $J/dF=4$ and the sum in Eq.~(\ref{3}) effectively contains 10 terms. Note that this figure depicts populations of the Wannier-Stark states. Populations of the lattice sites show additional oscillations with the Bloch frequency (see upper panel). 

\begin{figure}
\center
\includegraphics[width=8.5cm]{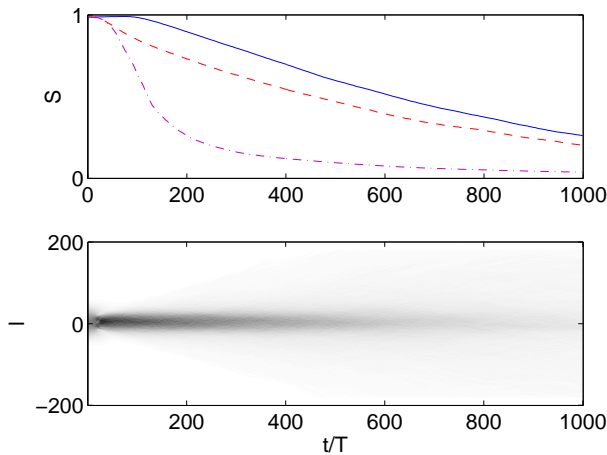}
\caption{Lower panel: Dynamics of the original system (\ref{1}) in the Wannier basis for  $\phi=\pi$, $\sigma=20$, $N=10^6$, $J=2$, $F=0.5$, $F_\omega=0.1F$, and  $g=10$. These parameters correspond to $\widetilde{J}=0.1$ and $\tilde{g}=100$. Upper panel: Survival probability for the coherent initial wave-packet (\ref{10}) and $\tilde{g}=100$ (solid line), for the incoherent initial wave-packet and $\tilde{g}=100$ (dashed line), and for the incoherent initial wave-packet and $\tilde{g}=0$ (dash-dotted line).}
\label{Fig4}
\end{figure}

The result depicted in Fig.~\ref{Fig3} refers to moderate nonlinearity $g=0.8$ or $\tilde{g}=g/\widetilde{J}=4$, where the system dynamics are stable. Larger atom-atom interactions lead to unstable dynamics and cause a rapid decoherence of the initial BEC state. Then, instead of soliton formation, one meets the phenomenon of temporal self-trapping. This phenomenon is exemplified in the lower panel in Fig.~\ref{Fig4}, which shows evolution of the site populations for $\tilde{g}=100$.  It is seen that atoms are temporally trapped by the nonlinearity. A  convenient characteristic of this process is the survival probability $S(t)$, which we define as the relative number of atoms staying in the region of support of the initial wave-packet ($|l|\le50$ for $\sigma_0=20$). The function $S(t)$ is depicted by the solid line in the upper panel in Fig.~\ref{Fig4}. It should be noted that $S(t)$ goes to zero for $t\rightarrow\infty$. This constitutes the main difference between the self-trapping in the DNLSE regime ($dF\gg J$) and in the regime where $dF<J$; in the former case $S(t)$ approaches a constant value defined by the relative number of permanently trapped atoms. 

Finally, we would like to stress one more time that the trapped atoms form an incoherent wave-packet, where the single-particle density matrix is close to a diagonal matrix.  In fact, for a systematic study of the incoherent self-trapping it is desirable to have an incoherent wave-packet from the very beginning. For the sake of comparison the dash-dotted and dashed lines in the upper panel in Fig.~\ref{Fig4} depict $S(t)$ for completely incoherent initial conditions and $\tilde{g}=0$ and $\tilde{g}=100$, respectively \cite{remark3}. In the former case the self-trapping is absent and the atoms spread ballistically \cite{preprint}. In the latter case, decay of the survival probability practically coincides with that depicted by the solid line if we take into account the time needed to decohere the initial BEC state.

{\em Conclusions.}
We have analyzed the dynamics of a BEC of cold atoms with repulsive interactions in driven tilted optical lattices.  It is shown that one can easily realize the focussing regime for the wave-packet evolution by properly adjusting the phase of the driving field with respect to the phase of Bloch oscillations. If the macroscopic interaction constant $g$ is larger than some critical value $g_{min}$, this focussing regime is followed by the formation of several bright solitons. We also checked that these solitons are stable against the interaction-induced decoherence  process present in the system. Further increase of the interaction constant above another critical value $g_{max}$ greatly enhances the decoherence process and, instead of soliton formation, one meets the phenomenon of incoherent self-trapping. In this regime a fraction of atoms is permanently or temporally trapped in an incoherent localized wave-packet, while the remaining atoms spread ballistically with the velocities defined by the effective hopping matrix element $\widetilde{J}$.

The discussed incoherent self-trapping has been recently observed in the laboratory experiment \cite{communic}. In the cited experiment the BEC of $N=1.2\times 10^5$ cesium atoms in vertical quasi one-dimentional lattice of the depth 3 recoil energies was subject to the sum of gravitational field and levitation force due to a magnetic field gradient, resulting in the Bloch frequency $\omega_B=2\pi\times 98$ Hz.  Modulating the magnetic field with the same frequency, the authors of Ref.~\cite{communic} realized the model (\ref{1}), where the 1D macroscopic interaction constant $g$ (which is proportional to the $s$-wave scattering length $a_s$) was varied in a wide interval by means of the Feshbach resonance. The regime of incoherent self-trapping was clearly observed at $a_s=336a_0$ ($a_0$ is Bohr radii), where the rate of spreading of the atomic cloud was essentially suppressed in constrast to the case $a_s=11a_0$. Simulating the dynamics of system (\ref{1}) for the specified parameters we concluded that to enter the regime of stable bright solitons, one should increase the number of atoms to $N\sim 10^6$ and further decrease the 1D interaction constant $g$. 

This work was supported by Russian Foundation for Basic Research. The author wishes to thank J. Brand for stimulating discussions and H.-C. N\"agerl and  E. Haller for sharing experimental results prior to publication.



\begin{thebibliography}{0}
\bibitem{remark0}
According to recent studies \cite{Grae09} the rate of decoherence due to atom-atom interactions is directly related to the Lyapunov exponents of the mean-field dynamics.
\bibitem{remark3}
The case of completely incoherent initial wave-packet is simulated by using ensemble of initial conditions with uniformly distributed phases of the amplitudes $a_l$.
\end{thebibliography}
\end{document}